\documentclass[aps,prb,twocolumn,showpacs,showkeys,floatfix]{revtex4}

\usepackage{graphicx,color}

\graphicspath{{figs/}}
\begin{document}
\title{Elastic Bending Modulus for Single-Layer Black Phosphorus}
\author{Hao-Yu Zhang}
    \affiliation{Shanghai Institute of Applied Mathematics and Mechanics, Shanghai Key Laboratory of Mechanics in Energy Engineering, Shanghai University, Shanghai 200072, People's Republic of China}
\author{Jin-Wu Jiang}
    \altaffiliation{Corresponding author: jwjiang5918@hotmail.com}
    \affiliation{Shanghai Institute of Applied Mathematics and Mechanics, Shanghai Key Laboratory of Mechanics in Energy Engineering, Shanghai University, Shanghai 200072, People's Republic of China}

\date{\today}
\begin{abstract}

We derive an analytic formula for the elastic bending modulus of single-layer black phosphorus (SLBP) based on the valence force field model. The obtained elastic bending modulus is 4.8028~{eV} and 7.9905~{eV} along the armchair and zigzag directions in the SLBP, respectively. These values are obviously larger than the bending modulus of 1.4~{eV} in graphene due to the intrinsic finite thickness for SLBP. Our derivation analytically illustrates that the elastic bending modulus of the SLBP is proportional to the square of the intrinsic thickness of the SLBP.

\end{abstract}

\maketitle
\pagebreak

\section{Introduction}
Few-layer black phosphorus (BP) is another interesting quasi two-dimensional system that has recently been explored as an alternative electronic material to graphene, boron nitride, and the transition metal dichalcogenides for transistor applications\cite{LiL2014,LiuH2014,BuscemaM2014,BuscemaM2014nc}. This initial excitement surrounding BP is because unlike graphene, BP has a direct bandgap that is layer-dependent.  Furthermore, BP also exhibits a carrier mobility that is larger than MoS$_{2}$\cite{LiuH2014}. The van der Waals effect in bulk BP was discussed by Appalakondaiah et.al.\cite{AppalakondaiahS2012prb} First-principles calculations show that single-layer BP (SLBP) has a band gap around 0.8~{eV}, and the band gap decreases with increasing thickness.\cite{DuY2010jap,LiuH2014} For SLBP, the band gap can be manipulated via mechanical strain in the direction normal to the BP plane, where a semiconductor-metal transition was observed.\cite{RodinAS2014,PengXH2014prb}

The single-layer BP has a characteristic puckered structure, which leads to the two anisotropic in-plane directions. As a result of this puckered configuration, anisotropy has been found in various properties for the single-layer BP, such as the optical properties,\cite{XiaF2014nc,TranV2014prb,LowT2014prb} the electrical conductance,\cite{FeiR2014nl} the mechanical properties,\cite{AppalakondaiahS2012prb,QiaoJ2014nc,JiangJW2014bpyoung,QinGarxiv14060261,WeiQ2014apl} and the Poisson's ratio.\cite{JiangJW2014bpnpr,QinGarxiv14060261,JiangJW2014bpsw}

The rippling phenomenon becomes unavoidable in low-dimensional materials, like the SLBP, as the elastic bending modulus is normally quite small for these one-atomic-thick structures. Graphene has very small bending modulus (around 1.4~eV),\cite{OuyangZC1997,TuZC2002,LuQ2009} so it is a highly flexible structure and its  properties can be manipulated through bending or bending induced strain. However, the bending phenomenon should be avoided in some graphene based electronic nano-devices. In such situation, the elastic graphene can be sandwiched by other two-dimensional materials with larger bending modulus,\cite{BritnellL2013sci} such as the MoS$_{2}$ with elastic bending modulus around 9.61~eV.\cite{JiangJW2013bend} Although the elastic bending modulus plays an important role for the two-dimensional material, the value of the elastic bending modulus for the SLBP has not been predicted to date, which is thus the focus of the present work.

In this paper, we analytically derive the elastic bending modulus for the SLBP, where the atomic interaction is described by the valence force field model (VFFM). The bending modulus is found to be 4.8028~{eV} and 7.9905~{eV} along the armchair and zigzag directions, respectively. The anisotropy in the bending modulus is discussed based on the geometrical coefficients in the SLBP. The large bending modulus as compared with graphene is attributed to the finite thickness of the SLBP, since the analytic derivation shows that the bending modulus is proportional to the square of the finite thickness.

\begin{figure*}[tb]
  \begin{center}
    \scalebox{1}[1]{\includegraphics[width=\textwidth]{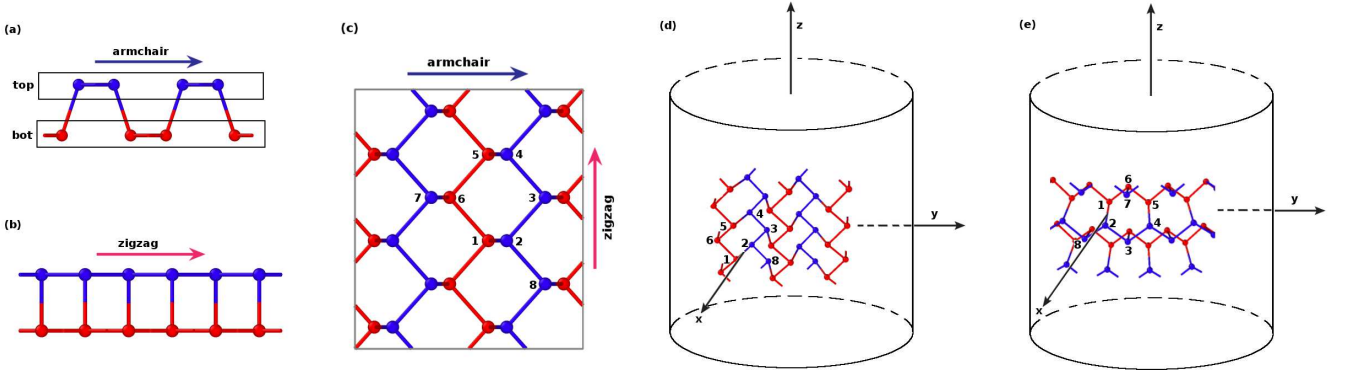}}
  \end{center}
  \caption{(Color online) Structure for SLBP. (a) and (b) are two side views. Atoms are divided into top group (blue online) and bottom group (red online). (c) Top view for the SLBP. (d) SLBP is bent by a curvature $\kappa$ along the armchair direction; i.e., SLBP is bent into a cylinder with radius $R=1/\kappa$. Atoms in the top group sit on the outer cylindrical surface with radius $R+d$, with $2d$ as the distance between the top and bottom groups. Atoms in the bottom group sit on the inner cylindrical surface with radius $R-d$. (e) SLBP is bent along the zigzag direction. The x-axis passes through the middle of atoms 1 and 2.}
  \label{fig_cfg}
\end{figure*}

\section{Structure and notations}
The atomic configuration of the SLBP is shown in Fig.~\ref{fig_cfg}~(a) for side view along the armchair direction and (b) for side view along the zigzag direction. Atoms are divided into the top group (blue online) and bottom group (red online). Fig.~\ref{fig_cfg}~(c) is the top view for the structure. Atoms 2, 3, 4, 7, and 8 are in the top group, while atoms 1, 5, and 6 are in the bottom group. The bond between atoms from the same group is the intra-group bond denoted by $d_1$, eg. $d_1=r_{23}=r_{16}$. The bond formed by atoms from different groups is the inter-group bond denoted by $d_2$, eg. $d_2=r_{12}$. In a similar way, $\theta_1$ is used to denote intra-group angles like $\theta_{328}$, which are formed by two intra-group bonds. We use $\theta_2$ to denote inter-group angles like $\theta_{321}$, which are formed by an intra-group bond and an inter-group bond.

We use the structure determined in the experiment.\cite{TakaoY1981physica} Two in-plane lattice constants are $a_1=r_{37}=4.376$~{\AA} and $a_2=r_{24}=3.314$~{\AA}. The out-of-plane lattice constant is $a_3=10.478$~{\AA}. There are four inequivalent atoms in the unit cell $\vec{a}_1\times\vec{a}_2$ of the SLBP, which will be chosen as atoms 1, 2, 3, and 6 in this work. The coordinate of these atoms are $\vec{r}_1=(-ua_1, 0, -va_3)$, $\vec{r}_2=(ua_1, 0, va_3)$, $\vec{r}_3=(0.5a_1-ua_1, 0.5a_2, va_3)$, and $\vec{r}_6=(-0.5a_1+ua_1, 0.5a_2, -va_3)$, where the origin of the Cartesian coordinate system is assumed to be the middle of $\vec{r}_{12}$. The x-axis is in the armchair direction and the y-axis is in the zigzag direction. The two dimensionless parameters are $u=0.0806$ and $v=0.1017$. The bond lengths from the experiment are $d_1=2.2449$~{\AA} and $d_2=2.2340$~{\AA}, and the two angles are $\theta_1=96.359^{\circ}$ and $\theta_2=102.094^{\circ}$.

A uniform bending is applied to the SLBP. More specifically, for a bending with curvature $\kappa$, the SLBP is rolled up onto a cylindrical surface with radius $R=1/\kappa$ along the armchair direction in Fig.~\ref{fig_cfg}~(d), and along the zigzag direction in Fig.~\ref{fig_cfg}~(e). After bending, the top atoms (blue online) sit on the outer cylindrical surface with radius $R+d$, and the bottom atoms (red online) are on the inner cylindrical surface with radius $R-d$, where $2d=2va_3$ is the distance between the top group and the bottom group.

It is necessary to introduce a subscript $\sigma=\pm$ to distinguish quantities in the outer and inner surfaces for the SLBP after bending. For instance, the intra-group bond length $r_{23}$ (from top group) and $r_{16}$ (from bottom group) has the same value of $d_1$ in the SLBP plane, but these two bonds have different value after bending. We introduce $d_1^{\pm}$ to denote intra-group bond length from the outer/inner cylindrical surface. The  subscript $\sigma$ is trivial for the inter-group bond length ($d_2^{\sigma}$) in the bent SLBP. However, we will keep this subscript in $d_2^{\sigma}$ in the following derivation for consistency. Similarly, we have the intra-group angle $\theta_1^{\sigma}$, with $\sigma=\pm$ for angles within the outer/inner surfaces. The angle $\theta_2^{+}$ is for the inter-group angle like $\theta_{321}$, whose peak (atom 2) is on the outer surface. The angle $\theta_2^-$ is for the inter-group angle like $\theta_{216}$, whose peak (atom 1) is on the inner cylindrical surface.

\section{Empirical energy density}

\begin{table*}
\caption{ Parameters (in eV\AA$^{-2}$ ) for the VFFM potential from Ref~\onlinecite{KanetaC1982ssc}.}
\label{tab_vffm}
\begin{tabular*}{\textwidth}{@{\extracolsep{\fill}}|c|c|c|c|c|c|c|c|c|}
\hline 
$K_{r}$ & $K_{r}'$ & $K_{\theta}$ & $K_{\theta}'$ & $K_{rr'}$ & $K_{rr'}'$ & $K_{r\theta}$ & $K_{r\theta}'$ & $K''_{r\theta}$\tabularnewline
\hline 
9.9715 & 9.4598 & 1.0764 & 0.9341 & 1.1057 & 1.1057 & 0.7207 & 0.7207 & 0.7207\tabularnewline
\hline 
\end{tabular*}
\end{table*}

Several empirical potentials have been developed to describe the atomic interaction for the SLBP, including the VFFM potential\cite{KanetaC1982ssc} and the Stillinger-Weber potential.\cite{JiangJW2013sw} Both potentials were fitted to the phonon dispersion of the SLBP. The Stillinger-Weber potential includes some nonlinear properties, so it can be applied in molecular dynamics simulations of the SLBP. The VFFM is a linear model, so it is suitable for the investigation of linear properties in the SLBP, like the elastic bending modulus studied in this work. The VFFM is convenient for deriving analytic expressions for elastic properties thanks to its simplicity. An analytic expression is of help for an explicit understanding of the elastic properties. Hence, we will apply the VFFM to derive an analytic formula for the elastic bending modulus of the SLBP.

There are nine terms in the VFFM potential,
\begin{eqnarray}
V_{r} & = & \frac{1}{2}K_{r}\left(\Delta d_{1}\right)^{2};\\
\label{eq_Vr}
V_{r}' & = & \frac{1}{2}K_{r}'\left(\Delta d_{2}\right)^{2};\\
V_{\theta} & = & \frac{1}{2}K_{\theta}d_{1}^{2}\left(\Delta\theta_{1}\right)^{2};\\
V_{\theta}' & = & \frac{1}{2}K_{\theta}'d_{1}d_{2}\left(\Delta\theta_{2}\right)^{2};\\
V_{rr'} & = & \frac{1}{2}K_{rr'}\left(\Delta d_{1}\right)\left(\Delta d_{1}\right);\\
V_{rr'}' & = & \frac{1}{2}K_{rr'}'\left(\Delta d_{1}\right)\left(\Delta d_{2}\right);\\
V_{r\theta} & = & \frac{1}{2}K_{r\theta}d_{1}\left(\Delta d_{1}\right)\left(\Delta\theta_{1}\right);\\
V_{r\theta}' & = & \frac{1}{2}K_{r\theta}'\sqrt{d_{1}d_{2}}\left(\Delta d_{1}\right)\left(\Delta\theta_{2}\right);\\
V''_{r\theta} & = & \frac{1}{2}K''_{r\theta}\sqrt{d_{1}d_{2}}\left(\Delta d_{2}\right)\left(\Delta\theta_{2}\right).
\label{eq_Vrtpp}
\end{eqnarray}
The VFFM describes the energy variation of the system due to a small change in the bond length ($\Delta b_i$) and the angle ($\Delta \theta_i$) with $i=1,2$, which are induced by bending in the present work. The $V_r$ term describes the bond stretching energy for intra-group bond lengths like $r_{23}$. The $V_r'$ term is the energy corresponding to the bond stretching for inter-group bond lengths like $r_{12}$. The $V_{\theta}$ term describes the energy associating with the variation of intra-group angles like $\theta_{234}$. The $V'_{\theta}$ term describes the energy variation due to the variation of the inter-group angles like $\theta_{123}$. The $V_{rr'}$ term describes the potential energy for the simultaneous variation of two different intra-group bonds like $r_{23}$ and $r_{24}$. The $V'_{rr'}$ term gives the potential energy for the simultaneous variation of bonds like $r_{21}$ and $r_{23}$. The $V_{r\theta}$ term is for the energy association with the simultaneous variation of an intra-group bond like $r_{32}$ and an intra-group angle like $\theta_{234}$. The $V'_{r\theta}$ term gives the potential energy for the simultaneous variation of an inter-group angle like $\theta_{123}$ and an intra-group bond like $r_{23}$.  The $V''_{r\theta}$ term gives the potential energy for the simultaneous variation of an inter-group angle like $\theta_{123}$ and an inter-group bond like $r_{12}$. All parameters are shown in Tab.~\ref{tab_vffm}. The unit of these parameters has been converted from dyne/cm in the original work to eV\AA$^{-2}$.

Based on the VFFM potential, the bending energy density ($W$) is
\begin{eqnarray}
W\times S_{0} & = & \sum_{\sigma=\pm}(2V_{r}+V_{r}'+2V_{\theta}+4V_{\theta}'+2V_{rr'}\nonumber\\
&&+4V_{rr'}'+4V_{r\theta}+4V_{r\theta}'+4V''_{r\theta}),
\label{eq_w}
\end{eqnarray}
where $S_{0}=a_{1}a_{2}$ is the area of the unit cell. The right-hand side gives the total bending energy for the unit cell $\vec{a}_1\times \vec{a}_2$.

\section{Bending modulus}

The bending modulus can be obtained through its definition,
\begin{eqnarray}
D & = & \frac{\partial^{2}W}{\partial\kappa^{2}}\nonumber\\
 & = & \frac{1}{S_{0}}\sum_{\sigma=\pm}(2\times K_{r}\alpha_{1}^{\sigma2}+\times K_{r}'\alpha_{2}^{+2}+2\times K_{\theta}d_{1}^{2}\beta_{1}^{\sigma2}\nonumber\\
 &  & +4\times K_{\theta}'d_{1}d_{2}\beta_{2}^{\sigma2}+2\times K_{rr'}\alpha_{1}^{+2}+4\times K_{rr'}'\alpha_{1}^{\sigma}\alpha_{2}^{\sigma}\nonumber\\
 &  & +4\times K_{r\theta}d_{1}\alpha_{1}^{\sigma}\beta_{1}^{\sigma}+4\times K_{r\theta}'\sqrt{d_{1}d_{2}}\alpha_{1}^{\sigma}\beta_{2}^{\sigma}\nonumber\\
&&+8\times K''_{r\theta}\sqrt{d_{1}d_{2}}\alpha_{2}^{\sigma}\beta_{2}^{\sigma}),
\label{eq_bending}
\end{eqnarray}
where the bending-induced variations for the bond length and the angle have been expressed as linear functions of curvature; i.e., $\Delta d_i^{\sigma}=\alpha_i^{\sigma}\kappa$ and $\Delta \theta_i^{\sigma}=\beta_i^{\sigma}\kappa$, with $i=1, 2$ and $\sigma=\pm$. We have introduced eight geometrical coefficients,
\begin{eqnarray}
\alpha_{1}^{\sigma} & = & \frac{\partial d_{1}^{\sigma}}{\partial\kappa}|_{\kappa=0};\\
\alpha_{2}^{\sigma} & = & \frac{\partial d_{2}^{\sigma}}{\partial\kappa}|_{\kappa=0};\\
\beta_{1}^{\sigma} & = & \frac{\partial\theta_{1}^{\sigma}}{\partial\kappa}|_{\kappa=0};\\
\beta_{2}^{\sigma} & = & \frac{\partial\theta_{2}^{\sigma}}{\partial\kappa}|_{\kappa=0}.
\end{eqnarray}
The elastic bending modulus is determined by these eight geometrical coefficients. We thus will derive analytic expressions for these eight geometrical coefficients in the following.

\subsection{Geometrical coefficients for armchair direction}

We first derive geometrical coefficients for the armchair direction. We examine the structure change for the SLBP after it is bent in the armchair direction as shown in Fig.~\ref{fig_cfg}~(d). For the intra-group bond length $d_{1}^{+}$ (eg. $r_{23}$), both atoms 2 and 3 are in the outer cylindrical surface with radius $R+d$. We find the lattice vector to be
\begin{eqnarray}
\vec{r}_{23} & = & \left(\begin{array}{c}
\left(R+d\right)\cos\left[\kappa\left(\frac{1}{2}-u\right)a_{1}\right]\\
\left(R+d\right)\sin\left[\kappa\left(\frac{1}{2}-u\right)a_{1}\right]\\
\frac{1}{2}
\end{array}\right)\nonumber\\
 & - & \left(\begin{array}{c}
\left(R+d\right)\cos\left(\kappa ua_{1}\right)\\
\left(R+d\right)\sin\left(\kappa ua_{1}\right)\\
0
\end{array}\right).
\end{eqnarray}
The first derivative of the lattice vector is
\begin{eqnarray}
\frac{\partial\vec{r}_{23}}{\partial\kappa}|_{\kappa=0} & = & \left(\begin{array}{c}
-\frac{1}{4}\left(\frac{1}{2}-2u\right)a_{1}^{2}\\
d\left(\frac{1}{2}-2u\right)a_{1}\\
0
\end{array}\right).
\end{eqnarray}
As a result, the first derivative of the bond length is
\begin{eqnarray}
\alpha_{1}^{+} & = & \frac{\partial r_{23}}{\partial\kappa}|_{\kappa=0}=\frac{1}{r_{23}}\vec{r}_{23}\cdot\frac{\partial\vec{r}_{23}}{\partial\kappa}\nonumber\\
&=&\frac{d}{d_{1}}\left(\frac{1}{2}-2u\right)^{2}a_{1}^{2}=1.0485\AA^{2}.
\end{eqnarray}
For the intra-group bond length $d_{1}^{-}$ (eg. $r_{16}$) on the inner cylindrical surface, an analogous derivation gives $\alpha_{1}^{-}=-\alpha_{1}^{+}=-1.0485\AA^{2}$. 

The inter-group bond length is $d_{2}^{+}=d_{2}^{-}=r_{12}$. The lattice vector $\vec{r}_{12}$ is
\begin{eqnarray}
\vec{r}_{12} & = & \left(\begin{array}{c}
2d\cos\left(\kappa ua_{1}\right)\\
2R\sin\left(\kappa ua_{1}\right)\\
0
\end{array}\right).
\end{eqnarray}
The first derivative of the lattice vector is
\begin{eqnarray}
\frac{\partial\vec{r}_{12}}{\partial\kappa}|_{\kappa=0} & = & 0.
\end{eqnarray}
The first derivative of the bond length is
\begin{eqnarray}
\alpha_{2}^{\pm} & = & \frac{\partial r_{12}}{\partial\kappa}=\frac{1}{r_{12}}\vec{r}_{12}\cdot\frac{\partial\vec{r}_{12}}{\partial\kappa}=0.
\end{eqnarray}

We now examine the variation for the intra-group angle $\Delta\theta_{1}^{+}$ (eg. $\theta_{328}$). From the definition $\cos\theta_{1}^{+}=\hat{n}_{23}\cdot\hat{n}_{28}$, we get the first derivative of the angle
\begin{eqnarray*}
\beta_{1}^{+} & = & \frac{\partial\theta_{1}^{+}}{\partial\kappa}|_{\kappa=0}=-\frac{1}{\sin\theta_{1}}\left(\hat{n}_{23}'\cdot\hat{n}_{28}+\hat{n}_{23}\cdot\hat{n}_{28}'\right)\\
 & = & -\frac{1}{\sin\theta_{1}}\left[2\frac{dd_{1}-\alpha_1}{d_{1}^{3}}\left(\frac{1}{2}-2u\right)^{2}a_{1}^{2}+\frac{2\alpha_1}{d_{1}^{3}}\frac{1}{4}a_{2}^{2}\right]\\
 & = & -1.0481\AA.
\end{eqnarray*}
For the other intra-group angle $\theta_{1}^{-}=\theta_{561}$, an analogous derivation gives $\beta_{1}^{-}=-\beta_{1}^{+}=1.0481\AA$.

For the inter-group angle $\theta_{2}^{+}$ (eg. $\theta_{321}$), we get the derivative of the angle,
\begin{eqnarray*}
\beta_{2}^{+} & = & \frac{\partial\theta_{2}}{\partial\kappa}|_{\kappa=0}\\
 & = & -\frac{1}{\sin\theta_{2}}\left[\frac{d}{d_{1}d_{2}}\left(\frac{1}{2}-2u\right)^{2}a_{1}^{2}+\frac{\alpha_1}{d_{1}^{2}d_{2}}2u\left(\frac{1}{2}-2u\right)a_{1}^{2}\right]\\
 & = & -0.5684\AA.
\end{eqnarray*}
For the other inter-group angle $\theta_{2}^{-}$ (eg. $\theta_{612}$), an analogous derivation gives $\beta_{2}^{-}=-\beta_{2}^{+}=0.5684\AA$.

\subsection{Geometrical coefficients for zigzag direction}

We now derive the geometrical coefficients for the zigzag direction; i.e., the SLBP is bent in the zigzag direction as shown in Fig.~\ref{fig_cfg}~(e). For the intra-group bond length $d_{1}^{+}$ (eg. $r_{23}$), the lattice vector in the bent SLBP is
\begin{eqnarray}
\vec{r}_{23} & = & \left(\begin{array}{c}
\left(R+d\right)\cos\left(\kappa\frac{1}{2}a_{2}\right)\\
\left(R+d\right)\sin\left(\kappa\frac{1}{2}a_{2}\right)\\
\left(-\frac{1}{2}+u\right)a_{1}
\end{array}\right)-\left(\begin{array}{c}
R+d\\
0\\
-ua_{1}
\end{array}\right).
\end{eqnarray}
The first derivative of the lattice vector is
\begin{eqnarray}
\frac{\partial\vec{r}_{23}}{\partial\kappa}|_{\kappa=0} & = & \left(\begin{array}{c}
-\frac{1}{8}a_{2}^{2}\\
\frac{d}{2}a_{2}\\
0
\end{array}\right).
\end{eqnarray}
The first derivative of the bond length is
\begin{eqnarray}
\alpha_{1}^{+} & = & \frac{\partial r_{23}}{\partial\kappa}|_{\kappa=0}=\frac{1}{r_{23}}\vec{r}_{23}\cdot\frac{\partial\vec{r}_{23}}{\partial\kappa}=\frac{1}{d_{1}}\frac{d}{4}a_{2}^{2}=1.3097\AA^{2}.\nonumber\\
\end{eqnarray}
For the other intra-group bond length $d_{1}^{-}$ (eg. $r_{16}$), an analogous derivation gives $\alpha_{1}^{-}=-\alpha_{1}^{+}=-1.3097\AA^{2}$.

The inter-group bond length is $d_{2}^{+}=d_{2}^{-}=r_{12}$. The lattice vector $\vec{r}_{12}$ is
\begin{eqnarray}
\vec{r}_{12} & = & \left(\begin{array}{c}
R-d\\
0\\
ua_{1}
\end{array}\right)-\left(\begin{array}{c}
R+d\\
0\\
-ua_{1}
\end{array}\right)=\left(\begin{array}{c}
-2d\\
0\\
2ua_{1}
\end{array}\right).
\end{eqnarray}
The first derivative of the lattice vector is
\begin{eqnarray}
\frac{\partial\vec{r}_{12}}{\partial\kappa}|_{\kappa=0} & = & 0.
\end{eqnarray}
The first derivative of the bond length is
\begin{eqnarray}
\alpha_{2}^{\pm} & = & \frac{\partial r_{12}}{\partial\kappa}=\frac{1}{r_{12}}\vec{r}_{12}\cdot\frac{\partial\vec{r}_{12}}{\partial\kappa}=0.
\end{eqnarray}

For the variation of the intra-group angle $\Delta\theta_{1}^{+}$ (eg. $\Delta\theta_{328}$), we get the first derivative of the angle
\begin{eqnarray}
\beta_{1}^{+} & = & \frac{\partial\theta_{1}^{+}}{\partial\kappa}|_{\kappa=0}\nonumber\\
 & = & \frac{-1}{\sin\theta_{1}}\left[-\frac{1}{2}\frac{d}{d_{1}^{2}}a_{2}^{2}-\frac{2\alpha_{1}}{d_{1}^{3}}\left[-\frac{1}{4}a_{2}^{2}+\left(-\frac{1}{2}+2u\right)^{2}a_{1}^{2}\right]\right]\nonumber\\
 & = & 1.0503\AA.
\end{eqnarray}
For the other intra-group angle $\theta_{1}^{-}$ (eg. $\theta_{561}$), an analogous derivation gives $\beta_{1}^{-}=-\beta_{1}^{+}=1.0503\AA$.

For the inter-group angle $\theta_{2}^{+}$ (eg. $\theta_{321}$), we get the first derivative of the anlge
\begin{eqnarray}
\beta_{2}^{+} & = & \frac{\partial\theta_{2}}{\partial\kappa}|_{\kappa=0}\nonumber\\
 & = & -\frac{1}{\sin\theta_{2}}\left[\frac{d}{d_{1}d_{2}}\frac{1}{4}a_{2}^{2}-\frac{\alpha_1}{d_{1}^{2}d_{2}}2u\left(-\frac{1}{2}+2u\right)a_{1}^{2}\right]\nonumber\\
 & = & -0.7217\AA.
\end{eqnarray}
Similarly, for the other inter-group angle $\theta_{2}^{-}$ (eg. $\theta_{612}$), an analogous derivation gives $\beta_{2}^{-}=-\beta_{2}^{+}=0.7217\AA$.

\subsection{Discussions on bending modulus}

\begin{table}
\caption{Geometrical coefficients for bent SLBP with curvature $\kappa$. The other four parameters are $\alpha_{1}^{-}=-\alpha_{1}^{+}$, $\alpha_{2}^{-}=-\alpha_{2}^{+}$, $\beta_{1}^{-}=-\beta_{1}^{+}$ and $\beta_{2}^{-}=-\beta_{2}^{+}$. The dimension is $\AA^{2}$ and $\AA$ for $\alpha$ and $\beta$, respectively.}
\label{tab_alpha}
\begin{tabular}{|c|c|c|c|c|}
\hline 
 & $\alpha_{1}^{+}$ & $\alpha_{2}^{+}$ & $\beta_{1}^{+}$ & $\beta_{2}^{+}$\tabularnewline
\hline 
\hline 
arm & 1.0485 & 0.0 & -1.0481 & -0.5684\tabularnewline
\hline 
zig & 1.3097 & 0.0 & 1.0503 & -0.7217\tabularnewline
\hline 
\end{tabular}
\end{table}

\begin{figure}[tb]
  \begin{center}
    \scalebox{1.0}[1.0]{\includegraphics[width=8cm]{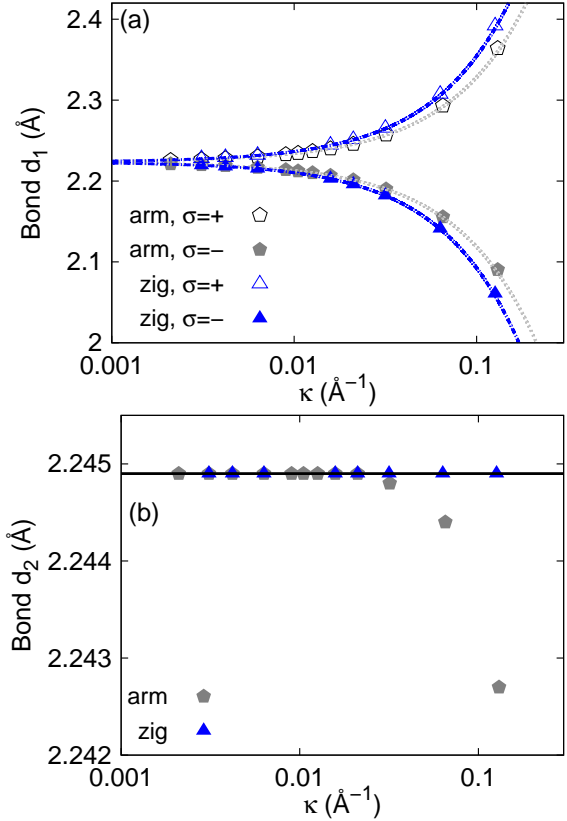}}
  \end{center}
  \caption{(Color online) The curvature ($\kappa$) dependence for bond length. The SLBP is bent either in the armchair or zigzag direction. (a) The intra-group bond length $d_{1}^{\pm}$. Numerical results are denoted by points. Lines denote the analytic expressions $ d_1^{\sigma}=d_1 + \alpha_1^{\sigma}\kappa$. (b) The inter-group bond length $d_2^{\pm}$. Numerical results are denoted by points. Lines denote the analytic expressions $ d_2^{\sigma}=d_2$.  Quantities with $\sigma=+$ have opposite behavior from those with $\sigma=-$.}
  \label{fig_alpha}
\end{figure}

\begin{figure}[tb]
  \begin{center}
    \scalebox{1.0}[1.0]{\includegraphics[width=8cm]{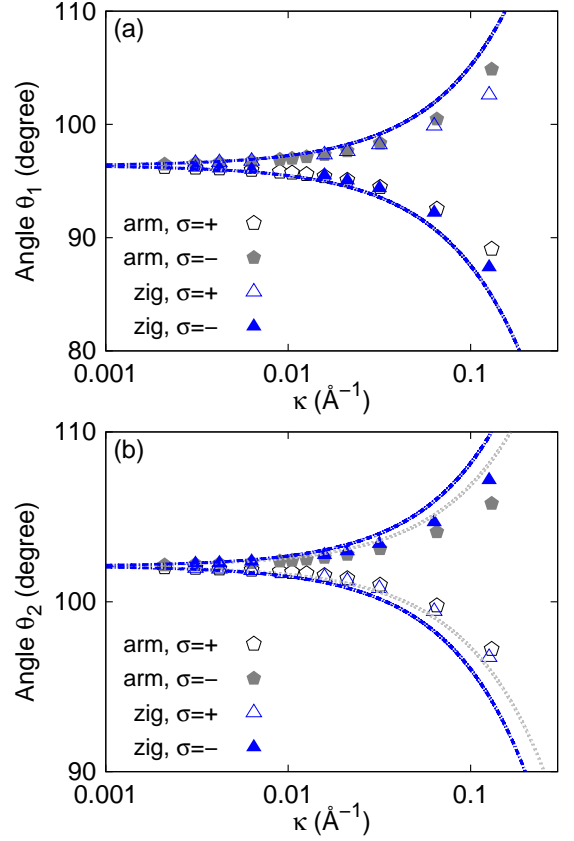}}
  \end{center}
  \caption{(Color online) The curvature ($\kappa$) dependence for angle. The SLBP is bent either in the armchair or zigzag direction. (a) The intra-group angle $\theta_{1}^{\pm}$. Numerical results are denoted by points. Lines denote the analytic expressions $ \theta_1^{\sigma}=\theta_1 + \beta_1^{\sigma}\kappa$. (b) The inter-group angle $\theta_2^{\pm}$. Numerical results are denoted by points. Lines denote the analytic expressions $ \theta_2^{\sigma}=\theta_2 + \beta_2^{\sigma}\kappa$.  Quantities with $\sigma=+$ have opposite behavior from those with $\sigma=-$.}
  \label{fig_beta}
\end{figure}

\begin{figure}[tb]
  \begin{center}
    \scalebox{1.0}[1.0]{\includegraphics[width=8cm]{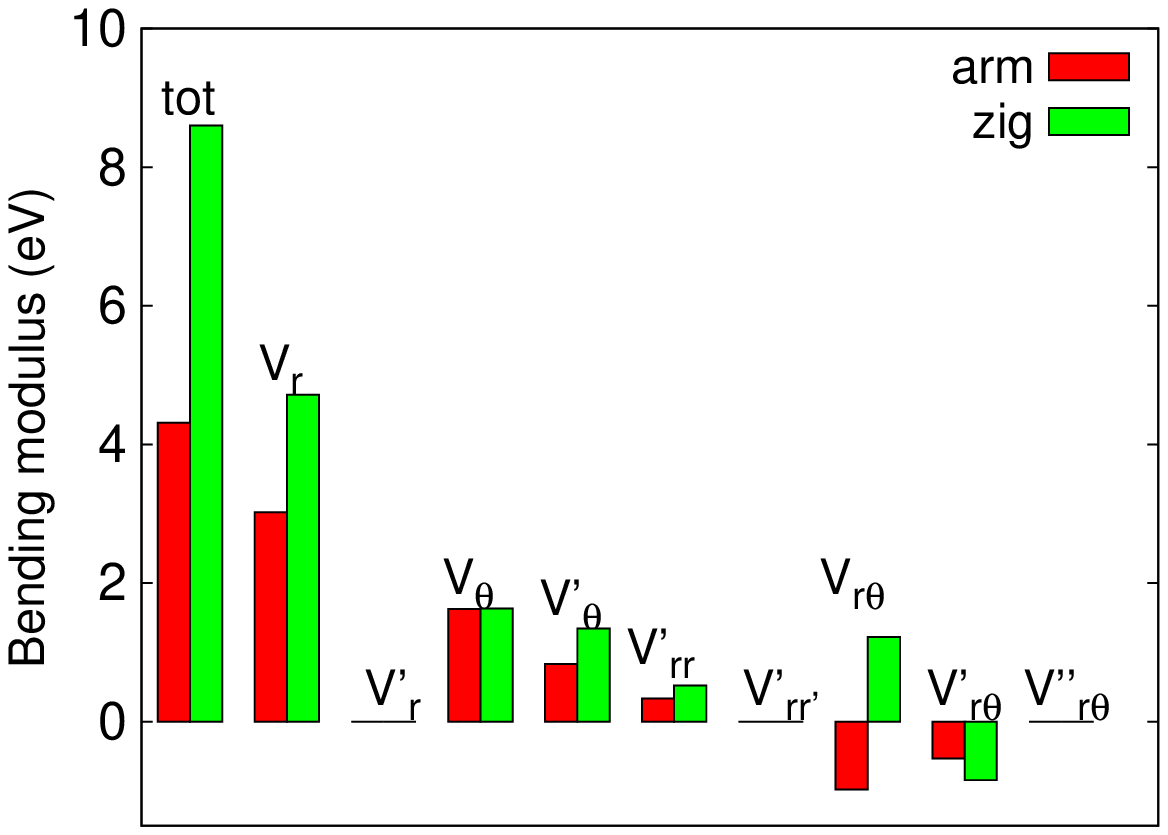}}
  \end{center}
  \caption{(Color online) The bending modulus in the SLBP that is bent in the armchair or zigzag direction. The bending modulus contributed by each VFFM potential term is displayed by a set of symbol. The total bending modulus is also displayed in the first set of symbol.}
  \label{fig_bending}
\end{figure}

In the above, all geometrical coefficients have been derived analytically for armchair and zigzag directions. These geometrical coefficients are summarized in Tab.~\ref{tab_alpha}. To validate the analytic derivations, these analytic results are checked against the numerical results in Figs.~\ref{fig_alpha} and~\ref{fig_beta}. Lines illustrate the analytic results, while numerical results are displayed by points. For a given curvature $\kappa$, the bond length $d_{i}^{\sigma}$ and angle $\theta_i^{\sigma}$ (with $i=1, 2$) are calculated numerically from a SLBP nanotube with radius $R=1/\kappa$, where P atoms are on the inner cylindrical surface with $R-d$ and outer cylindrical surface with $R+d$. Fig.~\ref{fig_alpha}~(a) shows that, the bending-induced variation for the intra-group bond $d_1^{\sigma}$ from the analytic expressions agree quite well with the numerical results up to $\kappa=0.1$~{\AA$^{-1}$}. There are some small deviations (less than 0.1\%) between the analytic results and the numerical results for the inter-group bond length $d_2$ for  $\kappa>0.06$~{\AA$^{-1}$} as shown in Fig.~\ref{fig_alpha}~(b). This deviation is quite small. Fig.~\ref{fig_beta} shows that the analytic expressions for the angle variation agree with the numerical results. These numerical calculations prove that the analytic expressions are valid and are suitable for bending with small curvature.

Inserting geometrical coefficients into Eq.~(\ref{eq_bending}), the obtained bending modulus is 4.3147~{eV} and 8.6014~{eV} along the armchair and zigzag directions, respectively. These values are sandwiched between the value of 1.4~{eV} for single-layer graphene\cite{OuyangZC1997,TuZC2002,LuQ2009} and the value of 9.61~eV for the single-layer MoS$_{2}$.\cite{JiangJW2013bend} Through the analytic derivation, we can find that the finite thickness is very important for the SLBP to have a higher bending modulus than graphene. More explicitly, all geometrical coefficients are proportional to the finite thickness ($2d$) for the SLBP. As a result, the bending modulus is proportional to $d^2$, which agrees with the well-known relationship from the shell theory, $D=E^{2D}h^2/(12(1-\nu^2))$, with $E^{2D}$ as the two-dimensional stiffness, $h$ as the thickness, and $\nu$ as the Poisson's ratio.

The bending modulus in the zigzag direction is obviously larger than the armchair direction. Fig.~\ref{fig_bending} compares the bending modulus value contributed by each VFFM potential term in the armchair and zigzag directions. All VFFM potential terms give larger bending modulus in the zigzag direction than the armchair direction, except the $V'_{r\theta}$ term that contributes larger value for the bending modulus in the armchair direction. There are three potential terms ($V'_r$, $V'_{rr'}$, and $V''_{r\theta}$) having no contribution to the bending modulus in both armchair and zigzag directions. It is because $\alpha_2^{\pm}=0$; i.e., the inter-group bond length $d_2$ doesn't change during the bending of the SLBP. It is quite interesting that $V_{r\theta}$ and $V'_{r\theta}$ contribute negative value for the bending modulus, because the bond length and angle have opposite variation behavior during the bending of the SLBP.

\section{conclusion}
To summarize, we have derived an analytic formula for the elastic bending modulus of the SLBP, using the VFFM potential. The obtained bending modulus is proportional to the finite thickness of the SLBP, which agrees with the elastic theory. The anisotropy in the bending modulus is discussed based on the individual contribution from each VFFM potential term. There are six VFFM potential terms that have nonzero contribution to the bending modulus, and only one of these six terms leads to larger bending modulus in the armchair direction, while the other five terms give larger bending modulus in the zigzag direction. As a result, the bending modulus is obviously larger in the zigzag direction.

\textbf{Acknowledgements} The work is supported by the Recruitment Program of Global Youth Experts of China and the start-up funding from Shanghai University.

%
\end{document}